\date{}
\documentstyle[aps,prl,twocolumn,epsf,epsfig,floats]{revtex}
\begin{document}
\draft
\twocolumn[\hsize\textwidth\columnwidth\hsize\csname @twocolumnfalse\endcsname

\title{Enhancement of  persistent currents  due to confinement in metallic samples}
\author{V. M. Apel, G. Chiappe and M. J. S\'anchez}

\address{\it Departamento de F\'{\i}sica  J. J. Giambiagi, Facultad de Ciencias Exactas  y Naturales,  
Universidad de Buenos Aires. Ciudad Universitaria, 1428 Buenos Aires, Argentina.} 

\maketitle

\begin{abstract}

Confinement and surface roughness (SR) effects on the magnitude of the persistent current 
are analized for ballistic bidimensional metallic samples. Depending on the particular geometry, localized
 border states can show up at half filling. These border
states contribute coherently to the persistent current and its magnitude 
is enhanced with respect to their value in the absent of confinement. A linear scaling of 
the typical current $I_{typ}$ with the number of conduction channels $M$ is obtained.
This result is robust with respect to changes in the relevant lengths of the
samples and to the SR. Possible links of our results to experiments are also discussed.  

\end{abstract}
]

In recent years, advances in  nanotechnology  made possible  to design mesoscopic samples  
in which the carriers are confined and mainly scattered by the boundaries of the system \cite{kouwen}. 
In this situation, where the elastic mean free path is much larger than the system size, the sample 
can be considered in the ballistic regime in order to compute the relevant transport quantities.
In addition, the potential that confines the carriers into the mesoscopic device is not
perfectly controlled in experiments. Therefore, even when the bulk disorder is absent, surface
roughness (SR) is present and as a consequence the carriers are irregularly scattered  
at the borders of the sample.

When the sample is threaded by a magnetic flux persistent currents  originate 
\cite{but,cheung1}.
Persistent currents (PC) have been studied \cite{cheung2,bouch,louis} in confined 
bidimensional cylindrical systems without SR employing discrete models
with periodic boundary conditions in one of the directions. 

Without reference to PC, a  discrete lattice has been considered recently in order to take
into account SR in metallic systems \cite{enri}. In the latter model, a kind of superficial
disorder was introduced through the absence of hopping to and from $L$ sites chosen at random 
among the $L^2$ sites defining the lattice.

In this letter  we study the PC for a model in which the confinement and the SR effects can be simultaneously analized. We consider circular (centred and non-centred) and square  bidimensional loops on a discrete square lattice. Therefore the  confinement breaks completely the translational  symmetry of the underlying lattice.
This introduces a qualitative difference with respect to cylindrically shaped systems, in which the translational symmetry is preserved in the azimuthal direction and the two degrees of freedom are decoupled. 
In addition, the samples studied by us have the geometry of those considered in the experiment of Ref.\cite{chandra}.

Although much work has been done concerning the influence of bulk disorder on persistent currents \cite{guhr},
less is known about the ballistic regime. Our goal is to characterize the effect of the 
confinement and the SR on the magnitude of the persistent current in clean metallic samples.
Confinement effects in the context of orbital magnetism were previously considered for 
semiconductor devices \cite{jala}.

The total current   can be calculated at zero temperature as 
$I \sim {\partial E /\partial \phi}$, with $E$ the total energy of the system and $\phi$ the magnetic flux. The value of $I_{typ} \equiv \sqrt{\int I^2 \; d\phi}$  depends on the number of open channels present in the system, $M$.
For free electron models with unbounded energy spectrum it is known that $I_{typ} \sim \sqrt{M}$ \cite{cheung2,jala}.
This behavior can be understood taking into account the properties of the crossings (or quasicrossings) that appear  in the spectrum when the flux is varied and frustrate the coherent increase of the current \cite{fs}.
As expected, the same scaling with $M$ holds  for tight binding models away from half filling \cite{bouch}.
At half filling and for  cylindrically shaped systems, it was found that very particular geometrical configurations of the sample could 
give rise to coherent contributions of all the channels, but those sharp configurations would be very difficult to obtain in an experimental set up \cite{cheung2,louis}.  

We consider a $N \times N$ cluster of the square lattice.
Taking a single  atomic level per  lattice site the Hamiltonian is,
$H =  \sum_{{\bf m}} \epsilon_{{\bf m}}|{\bf m}><{\bf m}| \; + \sum_{{\bf m;k}} t_{{\bf m,k}}|{\bf m}><{\bf k}|\;$, 
where $(m,n) \equiv \bf{m}$ labels the coordinates of the sites in the lattice.
All the lengths are given in units of the lattice constant $a$.
The hopping integrals
$t_{\bf{m};\bf{m'}}$ are restricted to nearest neighbours. Assuming that
the vector potential  has only azimuthal component we take $
t_{\bf{m};\bf{m+l}} = t \; \exp{( i \int_{\bf{m}}^{\bf{m+l}} \bf{A} {\bf{.}} d \bf{l} )}$, 
where the phase is measured in units of the quantum flux $\phi_{o}$ and 
$\bf{l}$ is a vector that
points from the site $\bf m$ to  any of its four nearest neighbours.
With the on-site energies, $\epsilon_{\bf m}$, 
we simulate the confining potential in order to get the required profile of the sample. 
We will first focus on the circular centred geometry. 
As the confining potential  and the 
underlying lattice have  different symmetry, SR is present in these samples
but the rotational symmetry in $\pi/2$ is preserved. 
The lattice parameter $a$ represents a cut off for the allowed wavelengths and therefore the SR will be relevant near half filling, that is in the metallic regime. 
The two  length scales are  chosen as the internal  radius  $r$ and the  width of the sample $W$. 
While the second determines $M$, the ratio $s \equiv W / r$ controls the strength of SR. For  $s >> 1$ and 
for carriers with a wavelength of the order of $a$, two succesive scattering events by  the inner and outer 
boundaries of the sample are almost uncorrelated. 
On the other hand, for $s <<1$ those  events are strongly correlated. Therefore the 
eigenstates corresponding to these particular wavelengths will evolve from generic extended ones to  strongly localized as the value of $s$ is reduced. 

\begin{figure}
\epsfxsize=3.in
\epsffile{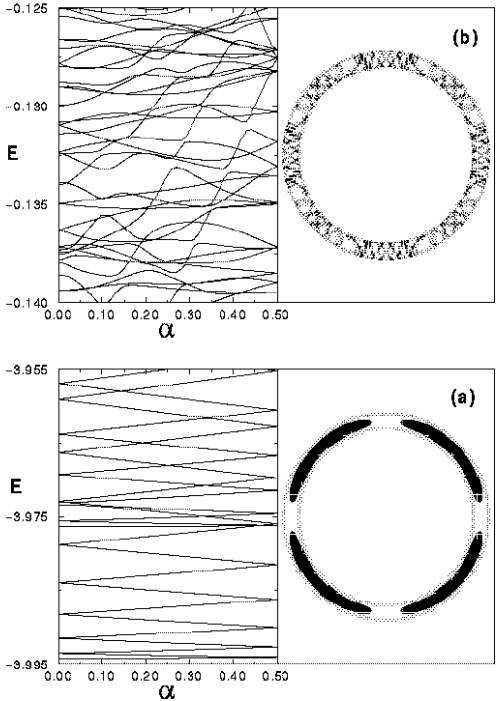}
\epsfxsize=3.in
\caption{(a) Energy levels at the bottom of the spectrum  as a function of the  
rescaled magnetic flux $\alpha$ for sample S1 together with a typical charge distribution. 
(b) Idem as (a) but for the quarter filling region of the spectrum.} 
\label{1}
\end{figure}

In Fig.~\ref{1} we show for sample $S1$ (with $W= 40$  and $r= 20$) the bottom (Fig.~\ref{1}(a)) and 
the quarter filling (Fig.~\ref{1}(b))  regions of the energy spectrum  as a function of the normalized 
flux $\alpha = \phi / \phi_{0}$ together with a generic  charge distribution for each region.
Figure~\ref{1}(a) corresponds to the largest wavelengths, that are the less sensitive  to the effect 
of the SR. This region of the spectrum looks qualitatively similar to the spectrum of an integrable 
Aharonov-Bohm annular billiard  \cite{fs}. 
Moreover, the generic  charge distribution corresponds to eigenstates with well defined value of the angular quantum number.

In the quarter filling  region the typical wavelengths are larger than $a$ but small  enough to make  the states sensitive to an effective polygonal boundary. This implies  the  lost of the continuous symmetry characteristic  of  the very large wavelengths. Therefore, as 
the energy rises towards  half filling,  avoided crossings show up.
The corresponding charge distributions are quite similar to those  for  a polygonal billiard and are extended states in general (see  Fig.~{1}(b)). 

\begin{figure}
\vspace{-1cm}
\epsfxsize=3.in
\epsffile{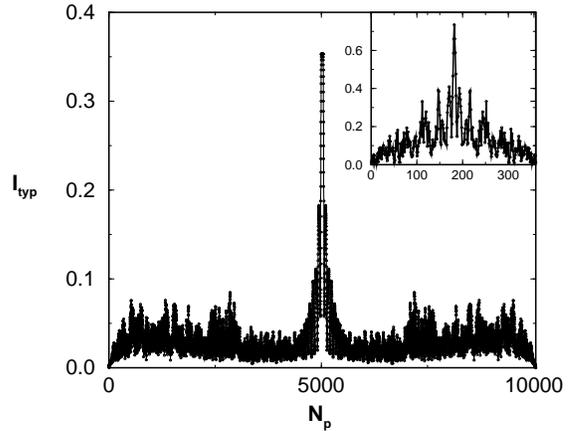}
\epsfysize=3.in
\vspace{0.5cm}
\caption{$I_{typ}$ as a function of $N_{p}$ for sample $S1$. Inset: Idem as (a) for a bidimensional square loop  with  outer square boundary of $20X20$ and inner one of $6X6$. Since the mean perimeter of the square loop doubles the mean perimeter of $S1$ the maximum value of $I_{typ}$ in the inset is almost twice
the value for $S1$.}  
\label{2}
\end{figure}

\begin{figure}
\epsfxsize=3.in
\epsffile{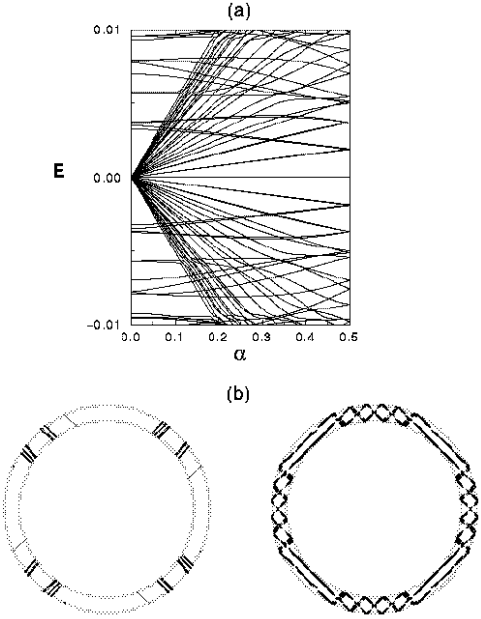}
\epsfysize=3.in
\caption{(a) Energy levels   close to half filling as a function of the rescaled magnetic flux $\alpha$ for $S1$.(b) Typical charge distributions for angularly localized states (left side) and radially localized states (right side). } 
\label{3}
\end{figure}

\begin{figure}
\epsfxsize=3.in
\epsffile{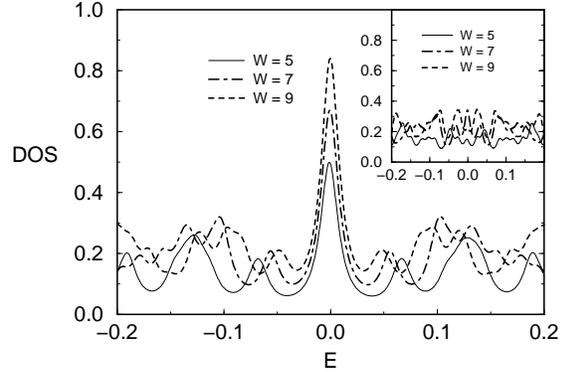}
\epsfysize=3.in
\caption{DOS for  square loops  with different widths $W$. Notice the peak  presents in the DOS at half filling. Inset: DOS for the rectangular samples. There is no bunching of states at zero energy in general.} 
\label{4} 
\end{figure}

In any case, when an occuppied level crosses (or quasicrosses)  an empty one, the total current $I$ exhibits
 a discontinuity (or abrupt oscillation) \cite{bouch}. 
As it is shown in Fig.~\ref{2}, below half filling, $I_{typ}$ is a 
highly fluctuating function of the number of particles, $N_{p}$. This is a consequence of the 
random distribution of crossings (or quasicrossings) that appears in these regions of the spectrum \cite{fs}.
The upper region of the spectrum, close to half filling, is shown in Fig.~\ref{3}(a) for $S1$.
In this region of energies the wavelengths are of the order of $a$.
A bunch of quasi-degenerate states appears at zero flux at half filling which  manifestes
as a peak in the density of states (DOS) at zero energy.
For other circular centred samples a qualitatively similar behavior was found.
In a bidimensional sample, a peak in the DOS at zero energy could appear for a very specific 
relation between the two relevant lengths of the sample. For example, for a cylinder of circumference $L$ 
(in units of $a$) and with $M$ transverse channels the 
condition is  $L = 2 \; (p / j) \; (M+1) \;$ \cite{cheung2}, where $p$ and $j$
are coprime integers.
For a rectangular sample, {\it i.e with Dirichlet boundary conditions in both directions}, with 
$L \times K$ sites the corresponding formula is $L + 1  = 2 \;
(p / j) \;(K+1) \;$.
As a consequence, for a square shaped sample the above relation is always 
satisfied and a peak in the DOS manifestes. 
We expect this result to remain valid in general for the
samples considered in this work, in which  Dirichlet boundary conditions 
are imposed on the inner and outer borders, providing the confinement
preserves the rotational  symmetry in $\pi /2$. This is confirmed by the
numerical results presented in  Figs.~\ref{4} for the square loops. Therefore, 
{\it independently of the number of channels in the loop}, the square and
centred circular geometries have a peak in the DOS at zero energy.
For the circular loops, as a consequence of the SR, the states that conform the 
peak in the DOS have different characteristics. Some of them are angularly 
localized states (see the  left side  of Fig.~\ref{3}(b)). They are not sensitive to the magnetic flux and correspond to the flat lines in the spectrum.
The other type of states are mostly radially localized and look like whispering
gallery modes. These states are originated by the
correlated scattering with the outer and inner boundary of the
sample that allows that the charge distribution does not fill completely the loop and the state remains localized. 
Althoug localized in one direction, these states go around the sample being
sensitive to the magnetic flux and carrying a finite current. Moreover, they
have a quasi well defined value of the angular quantum number that
characterizes the slope of the eigenenergies as a function of the flux.
The right side of fig.~\ref{3}(b) displays the 
characteristic charge distributions  for these states.

Opposite to what happens for systems with diagonal disorder, in the present model the  localization is 
obtained without introducing an additional scale of energy. Only remains the scale  defined by the 
kinetic energy. At zero flux, the energy of the localized states  will be very close to  zero  as  
a consequence of the symmetry and the finite character of the spectrum. 
Therefore, the existence of localized states in the system is manifested as a 
peak in the DOS that is equivalent to a finite gap in the spectrum of 
eigenergies. Before estimating  the magnitude of the gap we  remark that since the eigenenergies are degenerate  at $\alpha = 0$ 
but with quite different values of their slopes, each of the radially localized states must belong to 
a different conduction channel. 
Being radially localized, these states are quasi one dimensional and the energy gap is estimated as 
$\Delta_{g} = (\partial E / \partial k) \; \Delta k \gtrsim   ( \gamma \; A)^{-1} \;$, 
where $\Delta k$ is the inverse of the mean radius of the ring and $\gamma$ is a finite fraction of $W$
which measures the degree of spatial localization of the state. In order to check the above result we have done numerical calculations for samples with
same area $A$, obtaining the same value of $\Delta_{g}$.
The gap  has an important consequence on the values of $I_{typ}$  close to half filling. As it is shown 
in Fig.~\ref{2}, $I_{typ}$ increases monotonically with the number of occupied levels (number of particles) $N_{p}$. 

\begin{figure}
\epsfxsize=3.in
\epsffile{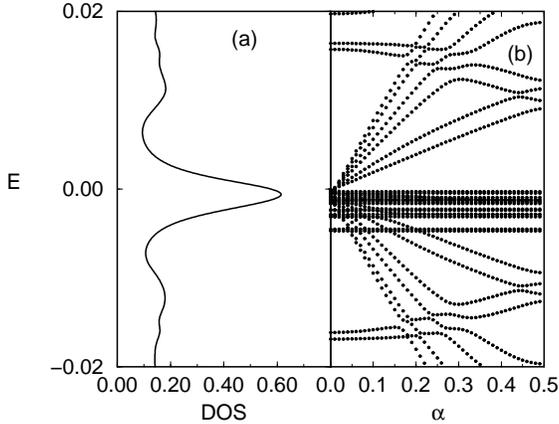}
\epsfysize=3.in
\caption{(a) DOS for a  non-centred circular loop of $W = 9$. (b) Energy
espectrum as function of the flux for the same sample. Flat lines at zero
energy  are associated with angulary localized states. Flux depending
eigenvalues begining at zero energy are associated with radially localized
states. The number of the latter (former) increases (decreases) with $W$.} 
\label{5} 
\end{figure}

Some comments must be made in connection  to the  non-centred circular loops.
Although in this case the SR does not preserve
the symmetry in  $\pi / 2 $, it affects the inner and outer boundary of the samples in the same way. This allows the correlated scattering between the inner and outer borders and therefore angularly and radially localized states also raise up at half filling, as it is
illustrated in the spectrum of Fig.~\ref{5}.
In the case of  the square loops, although the results are qualitatively similar to those obtained for 
the circular samples, some important differences must be noticed: $i)$ The square loops
do not have SR  and therefore there are not avoided crossings in the spectrum  as a function of the flux.
$ii)$ Border states carrying  a finite current were also found (analogous to the radially localized states of the circular samples). Therefore
$I_{typ}$ increases coherently near half filling as it is shown in the iinset of Fig.~\ref{2}. $iii)$ As a consequence of the absence of SR, angularly localized states do not exist at half filling (notice the absence of the plateau in $I_{typ}$ at half filling). 

For rectangular samples the localized border states  could
occur  for very specific relations between it lengths, and in general the 
spectrum has not bunching of  states at zero energy. In other words, 
a peak in the DOS is not expected at half filling for the rectangular geometry,
as it is shown in the inset of Fig.~\ref{4}.
\begin{figure}
\epsfxsize=3.in
\epsffile{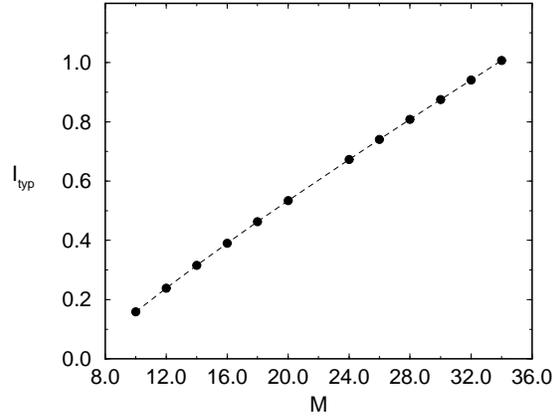}
\epsfysize=3.in
\caption{Linear scaling of $I_{typ}$ with the number of channels  $M$, at half
filling, for square loops with the same mean perimeter.}  
\label{6}
\end{figure}
We also find, in general, that when the localized  border states that carry a finite current  become dominant over the angularly localized states
 the scaling of $I_{typ}$ with $M$ at half filling is  substantially enhanced with respect to the  known scaling law out of half filling ($I_{typ}
\sim \sqrt{M}$). Fig.~\ref{6} shows the actual scaling law obtained for the square loops. 
Notice that the linear increase of $I_{typ}$ with $M$ is consistent with the fact that each of the localized states  bunched
at zero energy  belongs to a different conduction channel.  

The effect described in the present work can help  to understand some puzzles about PC in metallic rings. This is the case of the experiment of Ref.\cite{chandra} in 
which the PC of  isolated $Au$ rings 
was measured to be a factor $30 $ to $100$ times larger than the theoretical estimates.
In addition, in that experiment the reported value of the signal for the circular loop was $5$ times larger
than for the rectangular loop (although  the predicted theoretical values were roughly the same). 
Our results, even without providing a definitive explanation of the discrepancies between theory and experiment, are 
in line with the strong  enhancement of the current reported in it and provides the way to explain the observed differences between the circular and the
rectangular samples. In fact, as we previously described, for the circular loop we predict a peak in 
the DOS and therefore the enhancement of the PC. On the other hand, as we already discussed,
for the rectangular sample the enhancement is not expected in general. 
As a last point, we would like to remark that the enhancement of the current predicted in this work 
does  not have a critical dependence 
on the dimensions and on the SR of the samples,  making  their experimental realization feasible.
Unfortunately we are not aware of new measurements with other sample sizes in order to confirm or 
not our predictions. 
 
This work was partially supported by UBACYT (TW35-TX61), PICT97 03-00050-01015, CONICET and Fundaci\'on
Antorchas.  



\begin{thebibliography}{99}
\bibitem{kouwen} L.P. Kouwenhoven, C. M. Marcus {\it et.al.}, {\sl 
Proceedings of the NATO Advanced Study Institute on Mesoscopic Electron Transport}, edited by L.P. Kouwenhoven, L. Sohn and G. Sch\"on, Kluwer Series E, (1997).
\bibitem{but} M. B\"uttiker, Y. Imry and R. Landauer, {\sl Phys. Lett.} {\bf 96 A}, 365 (1983).
\bibitem{cheung1} H. F Cheung, Y. Gefen, E. K. Reidel, and Wei-Heng Shih,
 {\sl Phys. Rev. B} {\bf 37}, 6050 (1988).
\bibitem{cheung2} H. F Cheung, Y. Gefen and E. K. Reidel, {\sl IBM J. Res. Develop.}{\bf 32}, 359 (1988).
\bibitem{bouch} H. Bouchiat and G. Montambaux, {\sl J. Phys. France} {\bf 50}, 2695, (1989).
\bibitem{louis} E. Louis, J. A. Verg\'es and G. Chiappe, {\sl Phys. Rev. B} {\bf 11}, 6912, (1998). 
\bibitem{enri} J. A. Verg\'es and E. Louis, {\sl Phys. Rev. B} {\bf R3803} (1999).
\bibitem{chandra} V. Chandrasekar {\it et. al},  {\sl Phys. Rev. Lett.} {\bf 67}, 3578 (1991).
\bibitem{guhr}For a review see T. Guhr, A. M\"uller-Groeling and H. A. Weidenm\"uller, {\sl Physics Reports} {\bf 299}, 189 (1998).
\bibitem{jala}K. Richter, D. Ullmo and R. Jalabert, {\sl Phys. Rep.} {\bf 276}, (1996).
\bibitem{fs} A.J. Fendrik and M.J.S\'anchez, Eur. Phys. J. B, in press. cond-mat/9811272.
\end{thebibliography}
\end{document}